\documentclass[a4paper]{spie}  
\def\lapp{\ifmmode\stackrel{<}{_{\sim}}\else$\stackrel{<}{_{\sim}}$\fi}
\def\gapp{\ifmmode\stackrel{>}{_{\sim}}\else$\stackrel{>}{_{\sim}}$\fi}

\usepackage[]{graphicx}
\usepackage{color}

\newcommand{\AR}{\textcolor{black}}

\title{In-flight PSF calibration of the NuSTAR hard X-ray optics} 

\author{Hongjun An\supit{a}, Kristin K. Madsen\supit{b}, Niels J. Westergaard\supit{c},
Steven E. Boggs\supit{d}, Finn E. Christensen\supit{c},
William W. Craig\supit{d,e}, Charles J. Hailey\supit{f},
Fiona A. Harrison\supit{b}, Daniel K. Stern\supit{g}, William W. Zhang\supit{h}
\skiplinehalf
\supit{a}Department of Physics, McGill University, Montreal, Quebec, H3A 2T8, Canada; \\
\supit{b}Cahill Center for Astronomy and Astrophysics, California Institute of Technology, Pasadena, CA 91125, USA;\\
\supit{c}DTU Space, National Space Institute, Technical University of Denmark, Elektrovej 327, DK-2800 Lyngby, Denmark;\\
\supit{d}Space Sciences Laboratory, University of California, Berkeley, CA 94720, USA;\\
\supit{e}Lawrence Livermore National Laboratory, Livermore, CA 94550, USA;\\
\supit{f}Columbia Astrophysics Laboratory, Columbia University, New York NY 10027, USA;\\
\supit{g}Jet Propulsion Laboratory, California Institute of Technology, Pasadena, CA 91109, USA;\\
\supit{h}Goddard Space Flight Center, Greenbelt, MD 20771, USA}

\authorinfo{Further author information: (Send correspondence to H. An)\\H. An: E-mail: hjan@physics.mcgill.ca, Telephone: 1 514 398 6820\\}

\begin{document} 
  \maketitle 

\begin{abstract}
We present results of the point spread function (PSF) calibration of the hard X-ray optics
of the {\em Nuclear Spectroscopic Telescope Array (NuSTAR)}.
Immediately post-launch, {\em NuSTAR} has observed
bright point sources such as Cyg~X-1, Vela~X-1, and Her~X-1 for the PSF calibration.
We use the point source observations taken at several off-axis angles together with a ray-trace model
to characterize the in-orbit angular response, and find that the ray-trace
model alone does not fit the observed event distributions and applying empirical
corrections to the ray-trace model improves the fit significantly.
We describe the corrections applied to the ray-trace model and show that the uncertainties
in the enclosed energy fraction (EEF) of the new PSF model is
$\lapp$3\% for extraction apertures of $R\gapp60''$ with no significant
energy dependence. We also show that the PSF of the
{\em NuSTAR} optics has been stable over a period of $\sim$300 days during its in-orbit operation.
\end{abstract}



\medskip
\section{INTRODUCTION}

\label{sec:intro}  
\begin{figure}[tb]
\begin{center}
\begin{tabular}{cc}
\includegraphics[width=10cm]{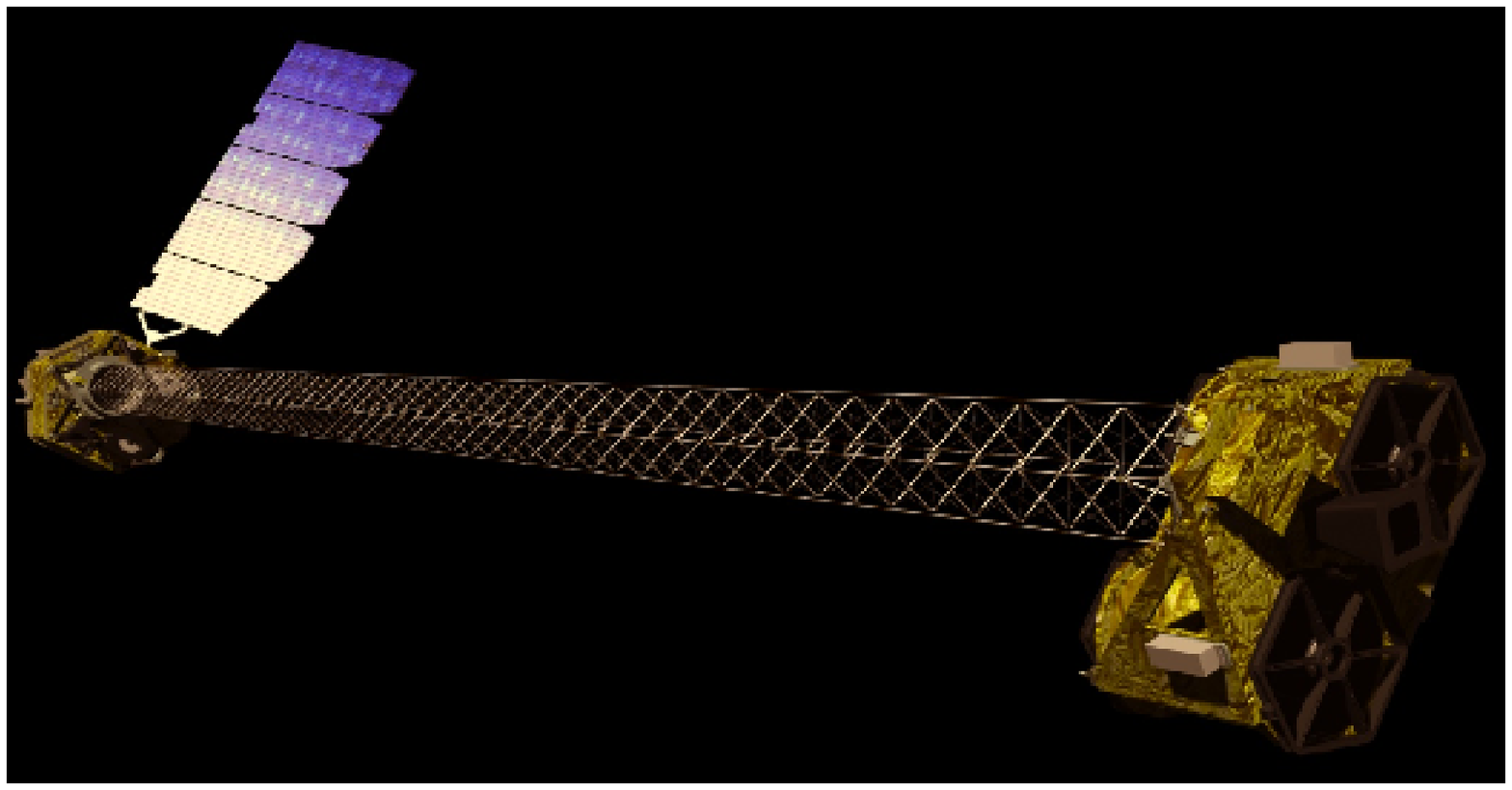} &
\includegraphics[width=6.4cm]{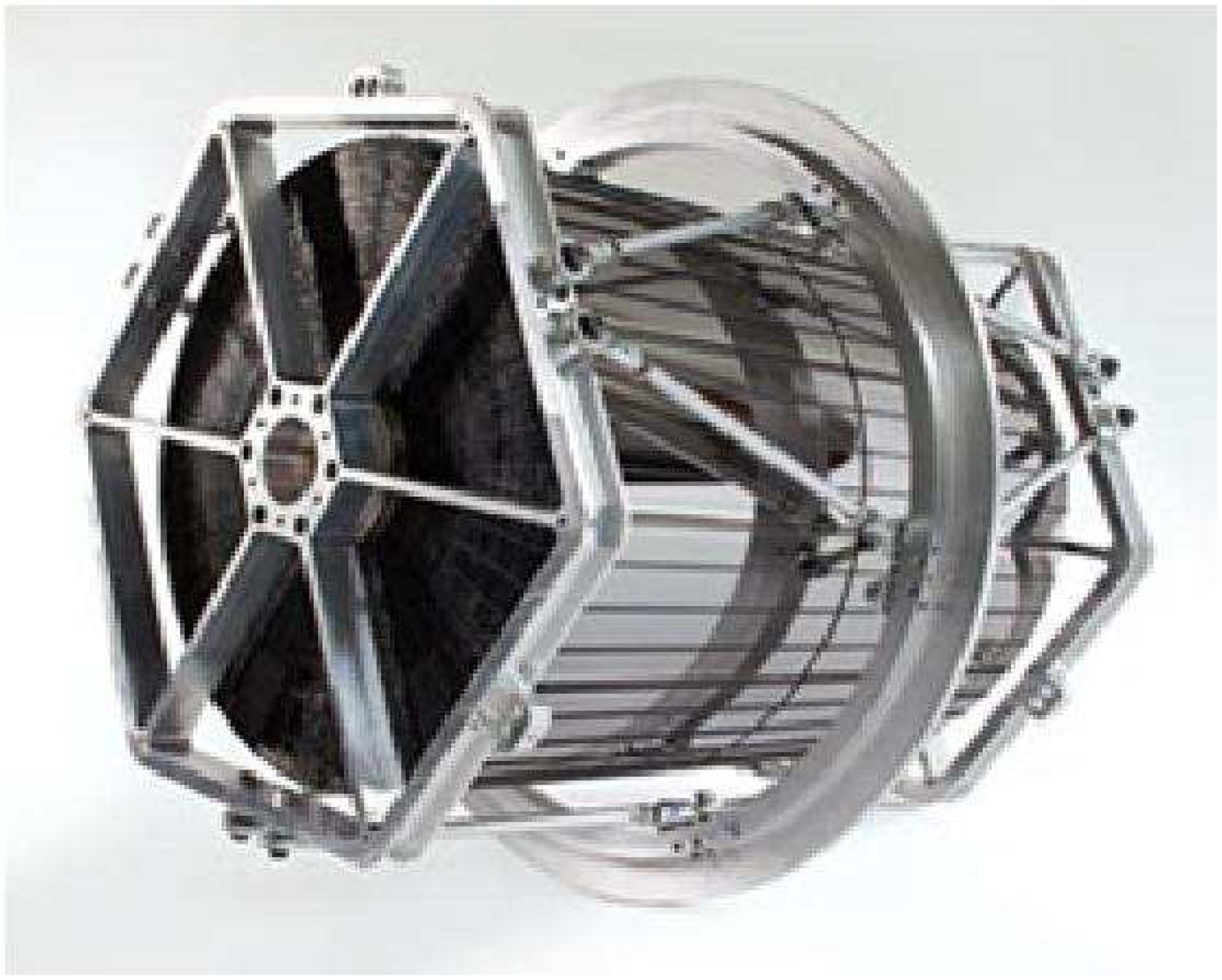} \\
\end{tabular}
\end{center}
\caption[example] 
{\label{fig:nustar} 
 The {\em NuSTAR} observatory (left) and the hard X-ray optic (right).
 The optic is $\sim$50 cm long and its focal length is $\sim$10 m.}
\end{figure}

The {\em Nuclear Spectroscopic Telescope Array (NuSTAR)} is the first hard X-ray focusing
telescope on orbit (Fig.~\ref{fig:nustar} left),
operating in the 3--79 keV band \cite{hcc+13, mhb+14}.
{\em NuSTAR} has two hard X-ray focusing optics and focal plane modules (FPMA and FPMB).
The hard X-ray optics employed in {\em NuSTAR} are segmented glass optics \cite{sjs+95} which are
composed of axially and azimuthally segmented mirrors stacked on a Titanium mandrel \cite{hab+10}.
Cylindrical mirrors are formed by thermally slumping flat glass onto a precision mandrel \cite{z09}.
The mirrors are coated with multilayer \cite{cjb+11}
and then bonded on precisely machined graphite spacers to form a conic approximation to the Wolter-I optics (Fig.~\ref{fig:nustar} right).
Each optic has 133 shells; the inner 90 shells are coated with Pt/C multilayer and the outer 43 shells with W/Si \cite{cab+11}.

The good angular response of the optics is crucially important not only to study extended sources such as
supernova remnants (SNRs) and pulsar wind nebulae (PWNe) but also to have better sensitivity for point source detection.
Furthermore, accurately measuring the enclosed energy fraction (EEF) with radius and energy is very important
since it not only determines the absolute normalization but may change the slope of Ancillary Response File (ARF).
PSF of the {\em NuSTAR} optics is mainly dominated by the glass figure error and alignment of the mirror layers.
Large scale figure errors ($\gapp$cm) in the mirror are efficiently suppressed during the mounting
process \AR{although it may introduce some large scale figure errors by forcing cylindrical mirrors to conical shapes.}
The small scale figure errors ($\lapp$cm) are hard to suppress and may be added as
distortion during mounting. The alignment between mirror shells was maintained by precisely machining
the spacers to the proper conical shape for each shell; any errors in the machining will cause image blurring.
During the optics assembly, the alignment was monitored, and the glass figures were measured.

Data for estimating the {\em NuSTAR} PSF on ground were taken with a laser metrology system, a mechanical probe,
and the X-ray calibration facility \cite{bcj+11}. The data were combined into a ray-trace model and
used to derive the PSF of the optics, and the Half Power Diameter (HPD) was estimated to be 52$''$ with an
uncertainty of $\sim$4$''$ \cite{kab+11}. The ground calibration was sufficient to ensure that the optics were
of the required quality, but because of the finite beam-line length and systematics in the measurements,
it is necessary to validate the PSF's in-orbit performance using bright point sources.
We focus on the PSF calibration of the {\em NuSTAR} optics in this paper, and more comprehensive
calibration of the observatory is presented elsewhere (Madsen et~al. 2014, in prep.).
We present the observations and data processes in Section~\ref{sec:obs}, and data analyses and the results
in Section~\ref{sec:dataana}, and then conclude in Section~\ref{sec:concl}.

\medskip
\section{Observations} 
\label{sec:obs}
We use bright point source observations to model the in-flight PSF.
The sources used for PSF calibration are listed in table~\ref{ta:psfobsid}.
The observations span a time interval of $\sim$300 days, and hence we can see how
and if the PSF of the optics changes over the interval.
The off-axis angle in an observation can change by $\sim1'$
because of the lack of pointing stability (e.g., relative motion between the optics and the detector benches) \cite{hcc+13},
and we report the average value of off-axis angle ($\theta$) distribution
of the observations in Table~\ref{ta:psfobsid}. Since the width of the off-axis angle distribution
is not large and the {\em NuSTAR} PSF does not rapidly change with off-axis angle,
we grouped the observations into two: one having the average off-axis angle $\lapp$2$'$
for the `near' on-axis PSF calibration and the other for the off-axis PSF.

We used observations 1 and 2 in Table~\ref{ta:psfobsid} for the
off-axis PSF calibration (Section~\ref{sec:offpsf}),
and all the rest for the `near' on-axis PSF (Section~\ref{sec:onpsf}),
and compared the event distribution of observation 5 with those of observations 7, 8, 10 and 11
in order to check the stability of the PSF over time (Section~\ref{sec:psftime}).
We processed the observations with {\ttfamily nupipeline} of NUSTARDAS v1.2.0 using standard
filtering procedure along with CALDB 20130509 to produce cleaned event files.
We further processed the cleaned event files for the analyses as described below.

\begin{table}
\caption{Observations used for PSF calibration}
\begin{center}       
\begin{tabular}{lcccccc}
\hline
Obs. \# & Source & Obs. ID & Obs. Date & Exposure & Off-axis angle$^{\rm a}$ & Comment \\
& & & (UTC) & (ks) & (arcminutes) \\
\hline
\hline
1 & Cyg~X-1		& 00001007001   & 2012-06-28  & 2.4 & 3 & For off-axis PSF\\
2 & Cyg~X-1		& 00001008001   & 2012-06-28  & 4.3 & 3 & For off-axis PSF\\
3 & GRS~1915$+$105	& 10002004001   & 2012-07-03  & 15  & 1 &  \\
4 & Cyg~X-1 	& 10002003001   & 2012-07-06  & 9   & 0.5 & \\
5 & Vela X-1	& 10002007001   & 2012-07-09  & 11  & 1 & \\
6 & GS0834		& 10002018001   & 2012-07-11  & 31  & 1 & \\
7 & Her~X-1		& 30002006002   & 2012-09-19  & 28  & 2 & \\
8 & Her~X-1		& 30002006005   & 2012-09-22  & 22  & 1 & \\
9 & Her~X-1		& 30002006007   & 2012-09-24  & 27  & 1 & \\
10 & Vela X-1	& 30002007002   & 2013-04-22  & 7   & 1 & \\
11 & Vela X-1	& 30002007003   & 2013-04-22  & 24  & 2 & \\
\hline
\end{tabular}
\end{center}
\footnotesize{$^{\rm a}${Average value for a distribution.\\}}
\label{ta:psfobsid}
\end{table}

\medskip
\section{Data analysis and Results} 
\label{sec:dataana}

\medskip
\subsection{Near On-axis PSF} 
\label{sec:onpsf}

\begin{figure}[t]
\centering
\begin{tabular}{cc}
\hspace{0.0 mm}
\includegraphics[width=3.38 in]{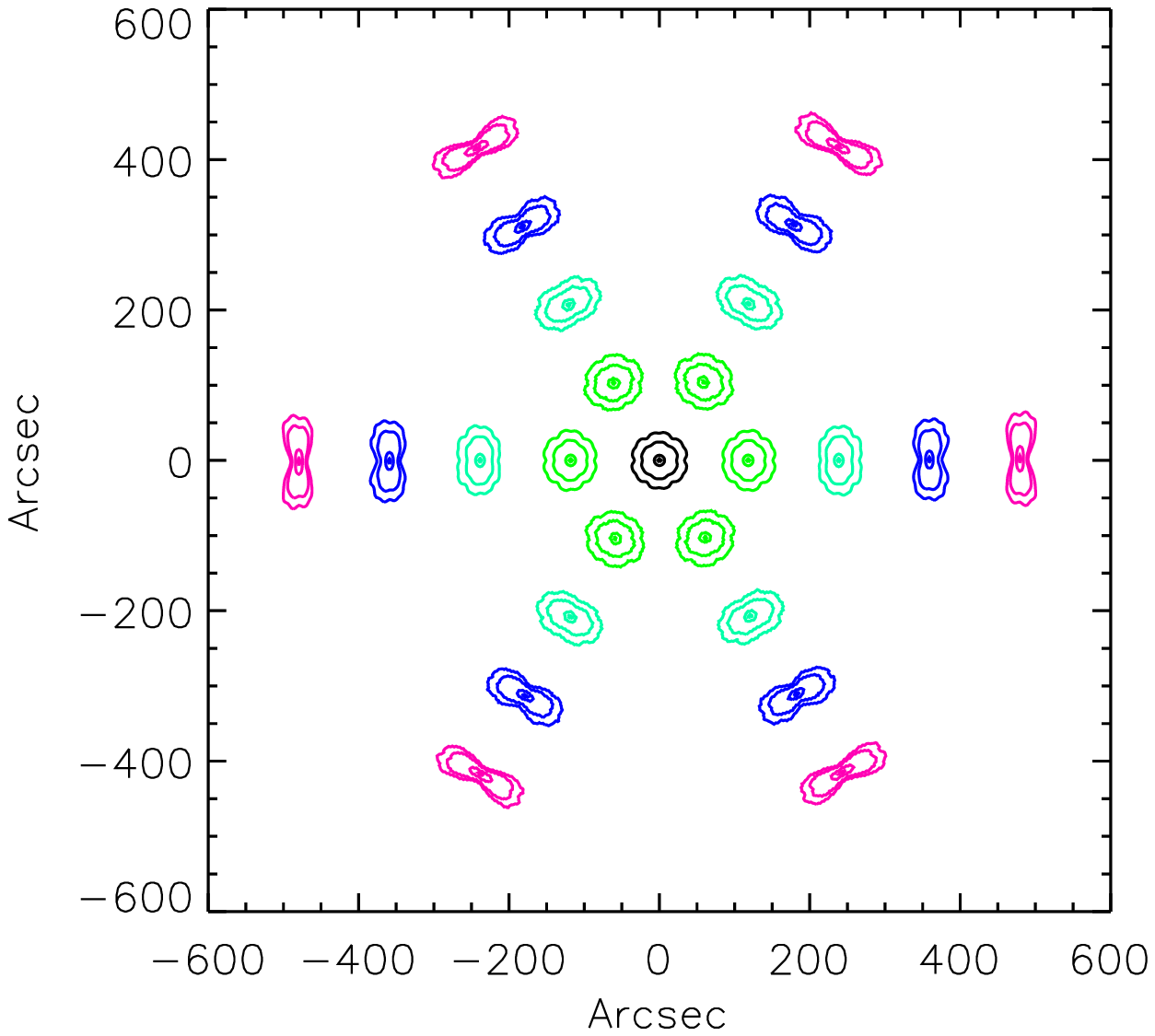} &
\hspace{-10.0 mm}
\includegraphics[width=2.47 in, angle=90]{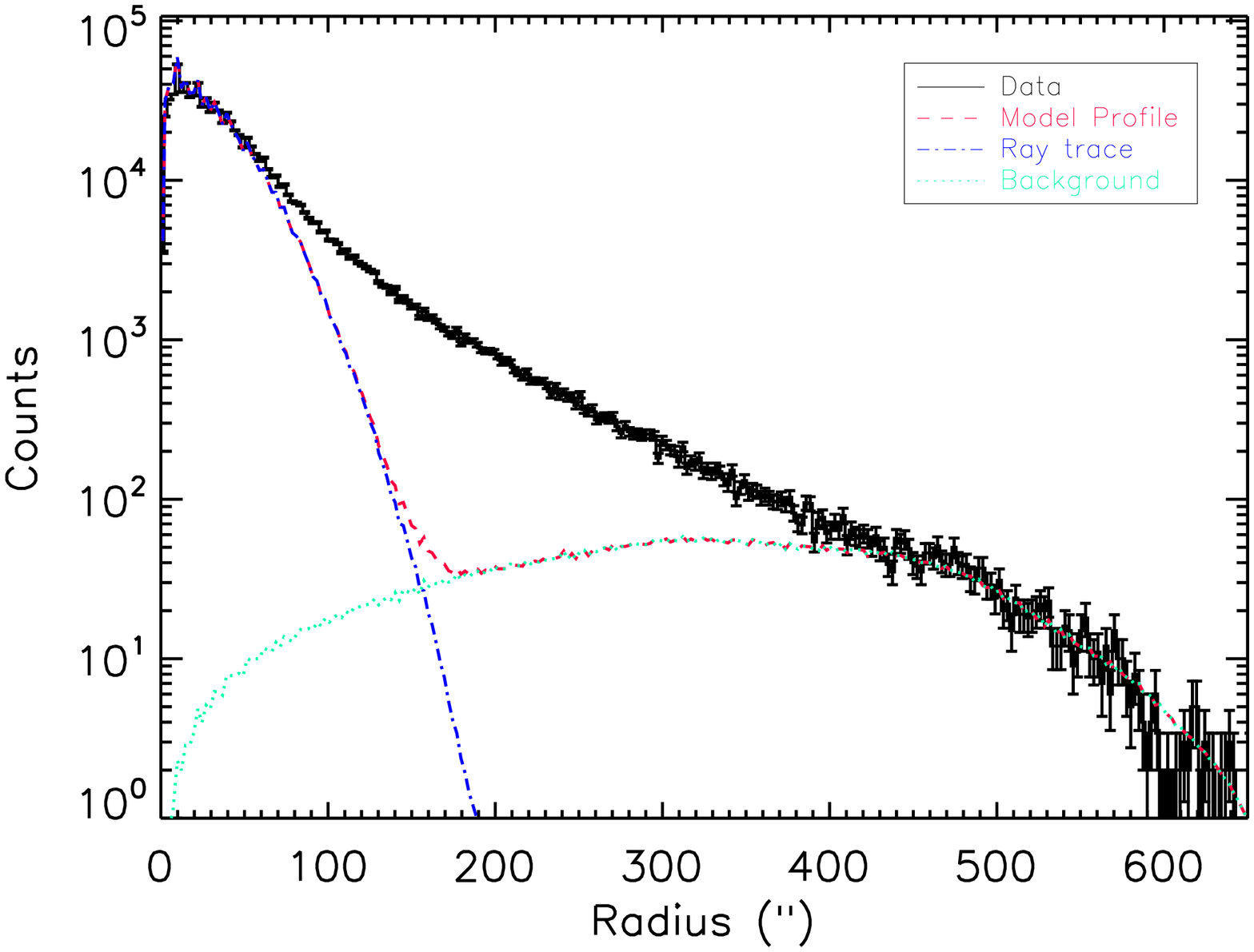} \\
\end{tabular}
\caption{{\it Left}: 2-D ray-trace PSF shapes at off-axis angles $\theta$ of 0$'$--8$'$ with a step of 2$'$.
Each color represents PSF for an off-axis angle. Off-axis PSFs were shifted to the
positive x-axis by the off-axis angle, and azimuthally rotated by 60$^{\circ}$--300$^{\circ}$.
{\it Right}: Radial profile of the Cyg~X-1 observation in the 5--8~keV band.
Best-fit model (red) was constructed with the ray-trace PSF (blue, Equation~\ref{eq:psf}) and background (cyan).
\label{fig:psffit}
}
\vspace{2mm}
\end{figure}

A ray-trace model was developed prior to launch \cite{w11}. The model uses the
surface figures of the mirrors measured with the mechanical probe scans after mounting \cite{cab+11}, and the
reflectivity measurements made with a sample of multilayer-coated mirrors \cite{cjb+11} as inputs, to trace
X-ray photons down to the focal plane detectors, producing the 2-D event distribution,
ray-traced PSF. We generated ray-traced PSFs at every 0.5$'$ out to 8.5$'$, shown in Figure~\ref{fig:psffit} left.
In the Figure, we shifted the PSF for an off-axis angle to the positive x-axis by the off-axis angle,
and azimuthally rotated the PSF ($\phi_{\rm off}$=0--300$^{\circ}$ with a step of 60$^{\circ}$) to show
the 2-D map. The Figure shows that the PSF is very circular out to $\theta \sim 2'$, and gradually distorts to a
bowtie shape at large off-axis angles.

For comparison with the ray-traced PSF, we extracted events from the point source observations (see Table~\ref{ta:psfobsid})
in several energy ranges (3--4.5 keV, 4.5--6 keV, 6--8 keV, 8--12 keV, 12--20 keV and 20--79 keV),
and produced observed radial profiles centered at the intensity peak for each observation.
Note that the ray-traced PSF does not have the energy dependence
\AR{because we do not have the ground measurements below $\sim$8 keV\cite{kab+11}.}
However, we resolved the event distribution in the energy space in order to see any change of the distribution with energy.
Since the off-axis angle was changing during an observation,
we weighted the ray-traced PSF with the aspect solution and vignetting function of the observation. The model PSF is formulated as 
\begin{equation}
\label{eq:psf}
\Psi(x,y)=\int \Psi_{\rm RT}(x-x_s^*,y-y_s^*,x_s(t), y_s(t)) \times \mathcal{V}(x_s(t), y_s(t)) dt
\times \mathcal{E}(x-x_s^*,y-y_s^*),
\end{equation}
where $\Psi(x,y)$ is the PSF model, $\Psi_{\rm RT}$ is the ray-traced PSF,
$x_s^*$ and $y_s^*$ are the sky positions of the source which are constant in time,
$x_s(t)$ and $y_s(t)$ are the source positions with respect to the optical axis
which determine the off-axis angle $\theta$ and $\phi_{\rm off}$,
$\mathcal{V}(x_s(t),y_s(t))$ is the vignetting factor for the off-axis angle, and
$\mathcal{E}(x-x_s^*,y-y_s^*)$ is the exposure map for the observation.
For source positions $x_s(t)$ and $y_s(t)$ at a time $t$, the ray-traced
PSF corresponding to $\theta$ is read in from the database and is rotated by $\phi_{\rm off}$ because
the ray-trace PSFs are stored for different $\theta$ values, but not
for different $\phi_{\rm off}$ in the database.
Note that the point source exposure is almost flat, and the overall shape of the PSF does not change
much by multiplying the exposure. However, it is important
for some localized effects such as detector gap.

\begin{figure}[t]
\centering
\begin{tabular}{ccc}
\hspace{-8.0 mm}
\includegraphics[width=1.8 in, angle=90]{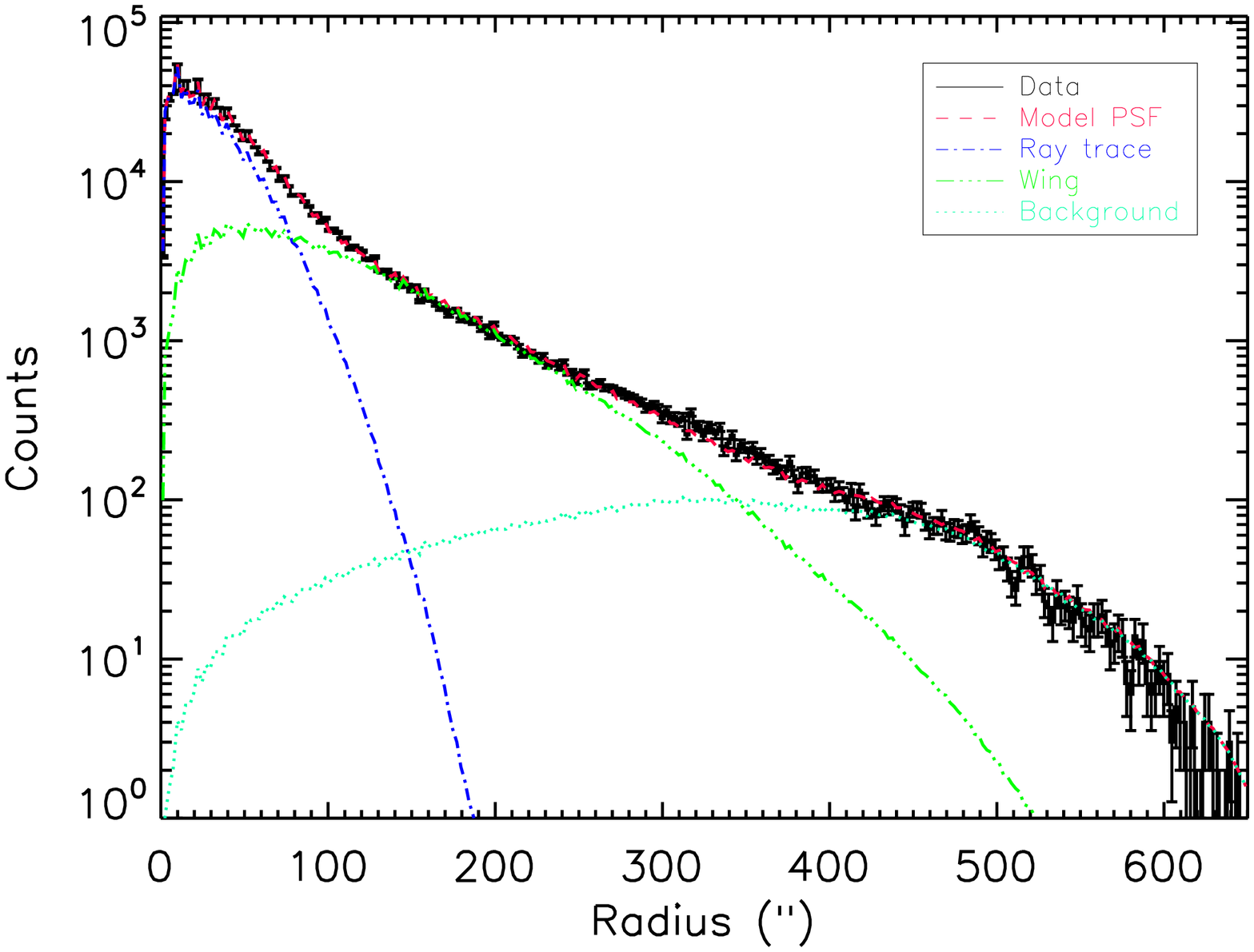} &
\hspace{-10.0 mm}
\includegraphics[width=1.8 in, angle=90]{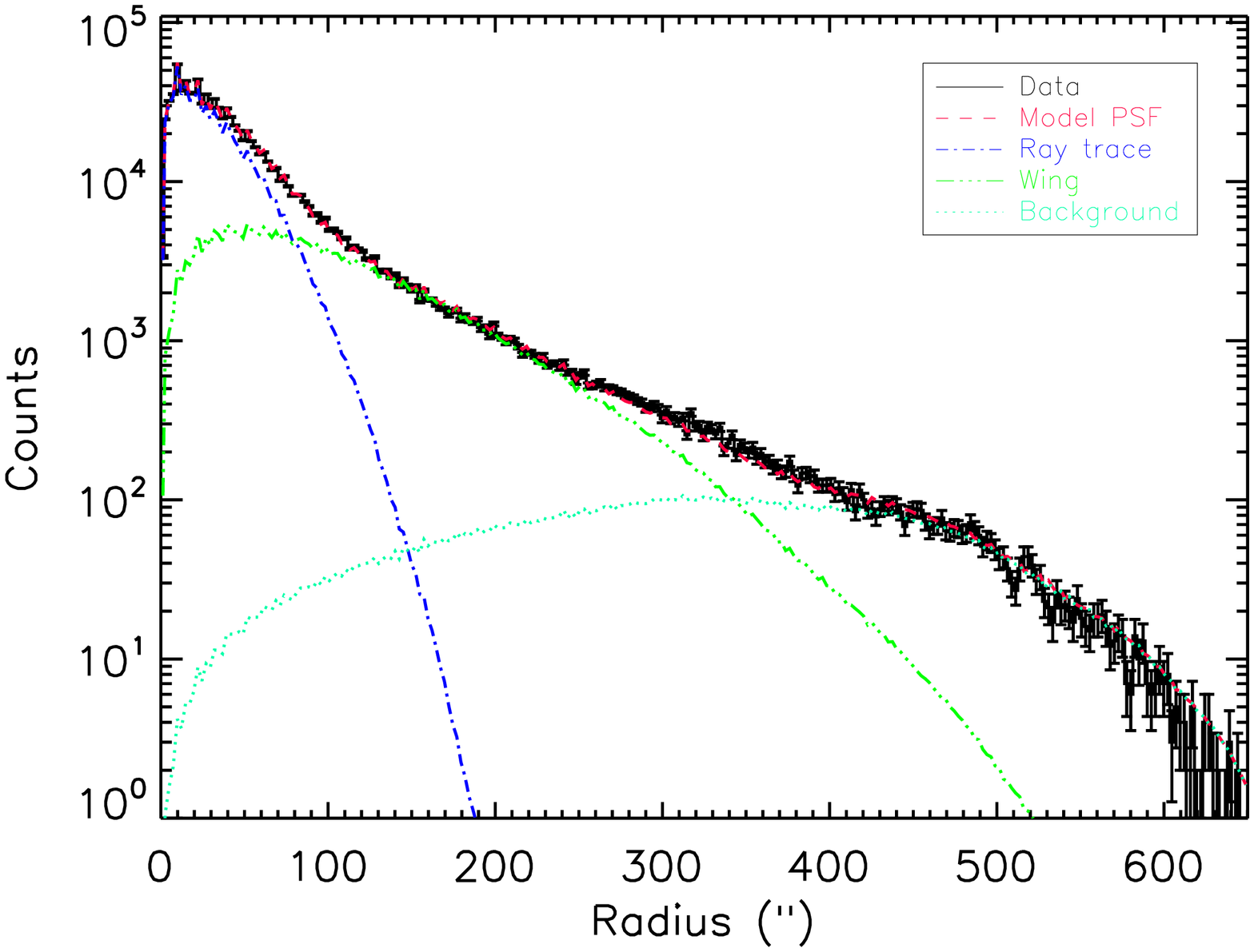} &
\hspace{-8.0 mm}
\includegraphics[width=1.8 in, angle=90]{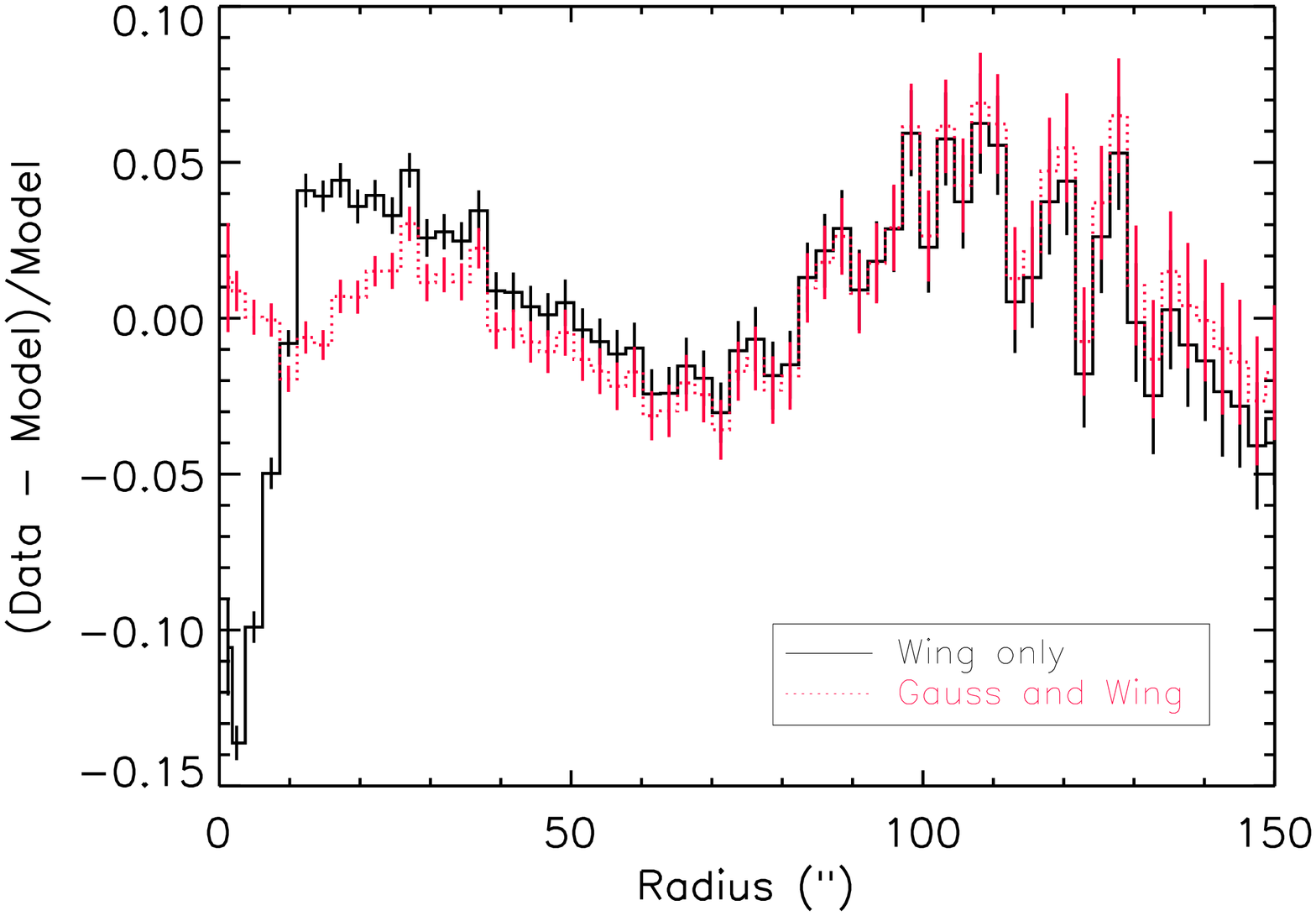} \\
\end{tabular}
\caption{{\it Left}: Radial profile (black) of the Cyg~X-1 observation in the 3--4.5~keV band and
the best-fit function. The best-fit function (red) is a combination of
the ray-trace PSF (blue, Equation~\ref{eq:psf}), the wing component (green, Equation~\ref{eq:psfw}),
and the background (cyan).
{\it Middle}: Same as the left panel but with a different fit function which is a combination of 
the ray-trace PSF with the Gaussian convolution (Equation~\ref{eq:psfG}),
the wing component (Equation~\ref{eq:psfw}), and the background.
{\it Right}: Residuals after the fits for the left panel (black) and the middle panel (red).
\label{fig:psfmodfit}
}
\vspace{2mm}
\end{figure}

Since the sources were very bright and we do not have a large enough region for background extraction in the detector,
the background was generated using a background model \cite{whm+14}.
The background model was constructed considering the internal, the aperture and
the cosmic X-ray background (CXB) components. The internal background is spatially uniform at low energies
but less uniform at high energies. The aperture component is nonuniform and mostly in the low energy band.
These two components do not go through the optics, and thus are unfocused. The CXB component goes through the optics
and is focused. The backgrounds were very small compared to the source, dominating only
above R$>$400$''$ (see Fig.~\ref{fig:psfmodfit}).

We fit the observed radial profiles with those of the model PSF in Equation~\ref{eq:psf} and the background,
adjusting the normalization constants only. An example of the fit is shown in Figure~\ref{fig:psffit} right.
After comparing with the near on-axis observations, we found that the PSF model in Equation~\ref{eq:psf} does not
describe the observed event distributions. Specifically, we find that
(i) there is a broad wing at $R$=100--400$''$ in the radial profiles of the observations,
(ii) the central cores of observed radial profiles are slightly broader than that of the model,
which cannot be explained with the current PSF model (see Fig.~\ref{fig:psffit}). We therefore modified
the PSF model.

In order to remove the large residuals in the wing ($R$=100--400$''$),
we added an exponential function to the PSF in Equation~\ref{eq:psf}.
The 2-D wing component is expressed in the following formula:
\begin{equation}
\label{eq:psfw}
\Psi_W(x,y)= \int e^{-a\sqrt{(x-x_s^*)^2 + (y-y_s^*)^2}}
\times \mathcal{V}(x_s(t), y_s(t))dt
\times \mathcal{E}(x-x_s^*,y-y_s^*).
\end{equation}
Note that the shape of the exponential function does not change with the off-axis angle at this point.
However, we further modify it below based on the off-axis observations (Section~\ref{sec:offpsf}).
For the core broadening, we used a Gaussian convolution model:
\begin{equation}
\label{eq:psfG}
\Psi_G(x,y)= \int \Psi_{\rm RT}(x-x_s^*,y-y_s^*,x_s(t), y_s(t)) \otimes G(x-x_s^*, y-y_s^*, \sigma)
\times \mathcal{V}(x_s(t), y_s(t)) dt
\times \mathcal{E}(x-x_s^*,y-y_s^*), 
\end{equation}
where $G(x-x_s^*, y-y_s^*, \sigma)$ is a Gaussian function with width $\sigma$. The modified PSF is
a combination of Equations~\ref{eq:psfw} and \ref{eq:psfG}.

We then added the background component, produced
the model radial profile, and fit the observed radial profile of each observation.
The fitting parameters are the amplitude for the ray-traced PSF ($C_{\rm 0}$),
width of the Gaussian convolution core ($\sigma$), amplitude and decay constant
for the exponential wing ($C_{\rm 1}$ and $a$), and amplitude for the background ($C_{\rm 2}$).
Note that we also included the off-axis correction (see Section~\ref{sec:offpsf}) in the PSF model.
Since fitting the observations simultaneously was not possible because they have different
aspect history and background, we calculated $\chi^2$ for
various values of fitting parameters for each observation, and found a set of parameters that
minimized the combined $\chi^2$.

An example the observed radial profile for
the Cyg X-1 observation (obs. 4) and the best-fit PSF model are shown
in Figure~\ref{fig:psfmodfit}. The wing correction in Equation~\ref{eq:psfw} compensates for the
large difference between the data and the model seen in Figure~\ref{fig:psffit},
but there are still large residuals near the core $R\lapp15''$
(see Figs.~\ref{fig:psfmodfit} left and right). The large residuals near the core are removed when applying the
core correction (Eq.~\ref{eq:psfG}, Fig.~\ref{fig:psfmodfit} middle and right).
We note that there are large scale residuals with amplitude of $\sim$5\%
even after applying all the corrections above (see the red curve in Fig.~\ref{fig:psfmodfit} right).
It was difficult to remove the large scale residuals for all the
nine observations with a common function because the residuals differ from observation to observation.
Therefore, we included the residuals in the error estimation for the EEF in Section~\ref{sec:psfresult}.

\medskip
\subsection{Off-axis PSF} 
\label{sec:offpsf}

The radial profile of the {\em NuSTAR} PSF does not change much with the off-axis angle, i.e.,
the FWHM remains roughly constant, but the 2-D shape of the PSF gradually distorts with increasing off-axis angle due to
geometrical shadowing of the shells, for example. As a result, the off-axis PSFs appear to be elongated
as shown in Figure \ref{fig:psffit} left, and a 2-D off-axis PSF model correction is required. 
Since accurately modeling the PSF at large off-axis angles is difficult due to the paucity of counts,
we matched the 2-D contours of the observations and the PSF
in order to obtain off-axis correction factors.

The 2-D distortion due to the shadowing is already considered in the ray-traced model. However, the
exponential wing is not yet incorporated into the ray-trace model, and thus we need to use a 2-D
analytic distortion model for the wing component.
Reduction in the effective area for a shell of a Wolter-I optic has been studied \cite{vc72,scbc+09}
and can be approximated with the following expression \cite{vc72}:
$$A_{\rm eff}(\theta)\approx A\left(1-\frac{2\theta}{3\alpha}\right) R(\alpha, \lambda),$$
where $\theta$ is the off-axis angle, $\alpha$ is the grazing incidence angle of on-axis photons, 
$R(\alpha,\lambda)$ is the reflectivity and $\lambda$ is the photon wavelength.
Note that azimuthal integration was performed in the above formula while we need the azimuth angle
dependence in order to have a 2-D shape.
Furthermore, we need to have an analytic expression for
the reflectivity, and integrate over the 133 layers of the {\em NuSTAR} optics,
which makes a detailed analytic implementation impossible.

Instead, we assumed that $A_{\rm eff}$ is a linear function of $\theta$, similar to the above,
and added the $\phi = \rm tan^{-1}((y-y_s^*)/(x-x_s^*))$
dependence as $\theta = \theta \rm cos^2(\phi)$, where we use $\rm cos^2(\phi)$ instead of
$\rm cos(\phi)$ for the $\phi$
dependence in order to avoid $A_{\rm eff}$ being greater than 1, the on-axis value,
in the left half plane (negative $x-x_s^*$).
Note that $\phi$ is calculated for each sky pixel $x$ and $y$, and is different
from the azimuthal angle of the source $\phi_{\rm off}=\rm atan^{-1}(y_{s}(t)/x_{s}(t))$.
In addition, we included an ellipticity factor in order to model the
elongation of PSF at off-axis angles along the direction transverse to the off-axis angle
(see Fig.~\ref{fig:psffit} left), and
used the following formulae to adjust the exponential wing at off-axis angles:
\begin{equation}
\label{eq:Awoff}
A_{\rm eff}(x,y,\theta(t))=1-A(\theta(t)){\rm cos^{2}}({\rm tan^{-1}}(B(\theta(t))(y-y_s^*)/(x-x_s^*))),
\end{equation}
and
\begin{equation}
\label{eq:psfwoff}
F_{\rm wing}(x,y,\theta(t),\phi_{\rm off}(t))=\mathcal R(\phi_{\rm off}(t)) 
\left(e^{-a\sqrt{(x-x_s^*)^2 + B(\theta)(y-y_s^*)^2}}\times A_{\rm eff}(x,y, \theta(t))\right),
\end{equation}
where $A(\theta)$ is a linear function which is 0 when $\theta=0$, $B(\theta)$ is the
ellipticity factor which is 1 when $\theta=0$, having the largest $A_{\rm eff}$ and no elongation for the on-axis PSF,
and $R(\phi_{\rm off}(t))$ is a rotation of a function around
the axis perpendicular to the $x$-$y$ plane centered at $x_s^*$ and $y_s^*$.
We did not apply the off-axis corrections to the Gaussian convolution model because the
observed event distribution matches well with the PSF at the core without the corrections.

\begin{figure*}[t]
\centering
\begin{tabular}{cc}
\hspace{0.0 mm}
\includegraphics[width=3.3 in, angle=90]{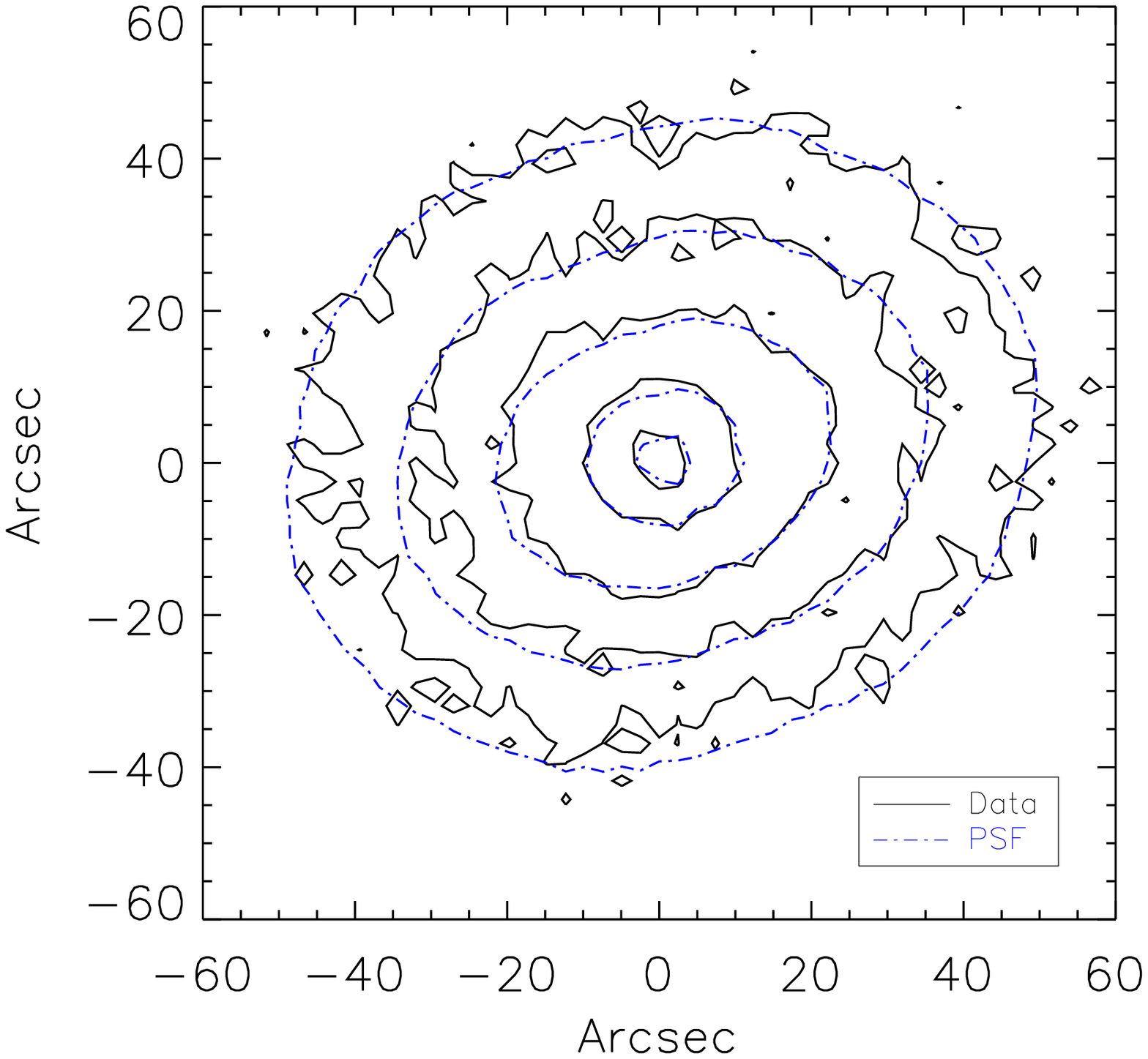}
\hspace{-24.0 mm}
\includegraphics[width=3.2 in, angle=90]{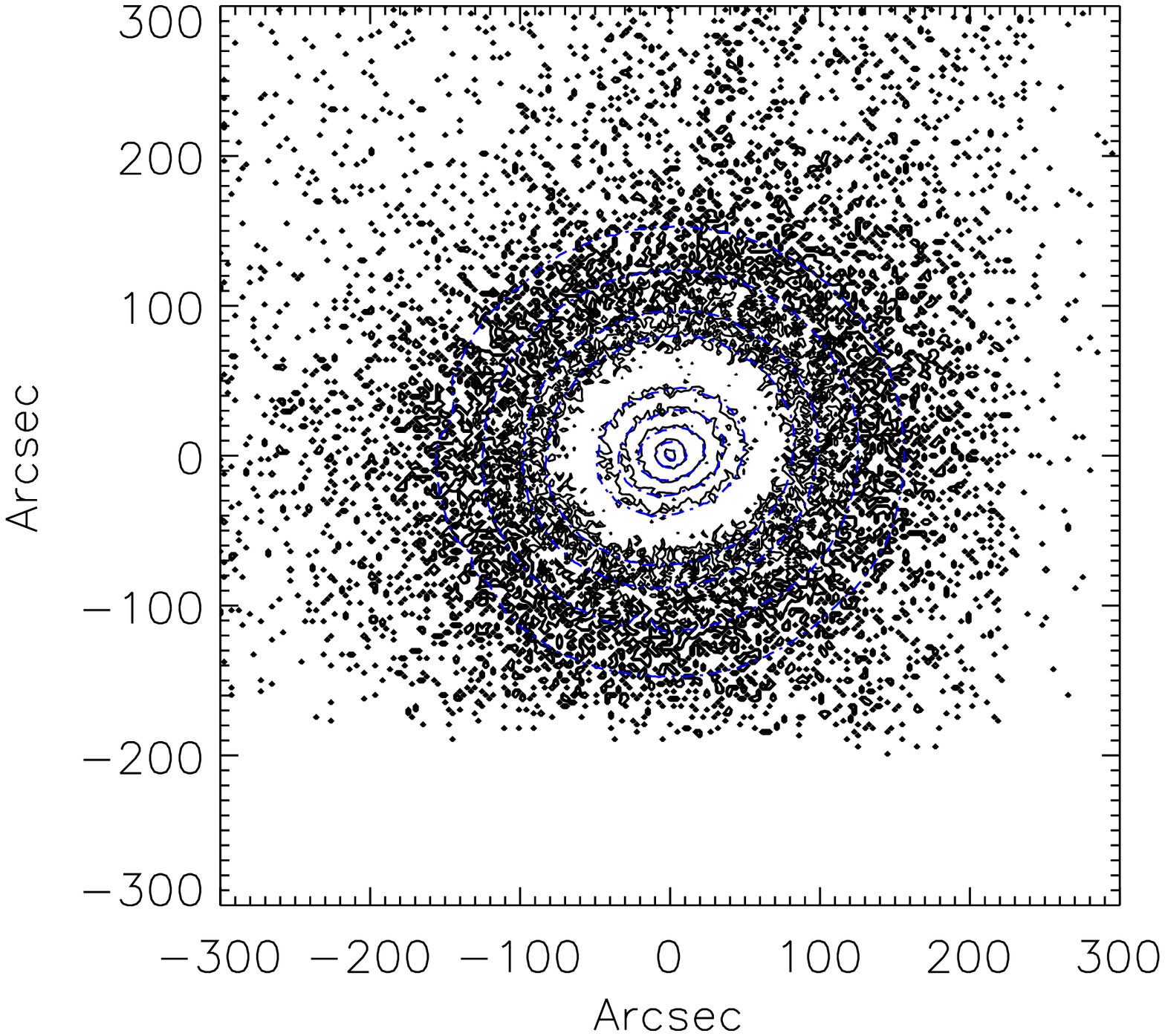} &
\end{tabular}
\vspace{-1.0 mm}
\caption{Count contours of observation 1 (black) and the model PSF (blue) for an off-axis angle $\theta \sim$3$'$
in a $2'\times2'$ field (left) and in a $10'\times10'$ field (right).
\label{fig:off}
}
\vspace{2mm}
\end{figure*}

The modified model PSF is a combination of Equations~\ref{eq:psfw} and \ref{eq:psfG}, but
replacing the exponential function in Equation~\ref{eq:psfw} with $F_{\rm wing}$ (Eq.~\ref{eq:psfwoff}).
We compared the model PSF with the off-axis observations ($R>3'$) listed in Table \ref{ta:psfobsid}.
We note that the ray-trace PSF dominates over the wing component in the inner part ($R\lapp 100''$)
of the 2-D distribution, and we had to use the outer part (R$>$100$''$)  where the wing component dominates
in order to determine the two functions $A(\theta)$ and $B(\theta)$. Since there are not enough events above
$R\sim100''$, we did not attempt to fit the data. Instead, we find linear functions $A(\theta)$ and $B(\theta)$
by matching the 2-D contours of the observations with that of the model PSF.
We find that the ray-traced PSF with the Gaussian convolution reproduces the observed 2-D distributions well
at smaller radii (Fig.~\ref{fig:off} left)
and that $A(\theta)=0.025\theta$ and $B(\theta)=1-0.025\theta$ made the contours match
at large radii (see Fig.~\ref{fig:off} right). We note that impact of including the off-axis effect
($A(\theta)$ and $B(\theta)$) is small for the near on-axis PSF.

\medskip
\subsection{Results of PSF calibration of the {\em NuSTAR} Optics}
\label{sec:psfresult}
\begin{figure}[t]
\centering
\begin{tabular}{cc}
\hspace{10.0 mm}
\includegraphics[width=2.45 in, angle=90]{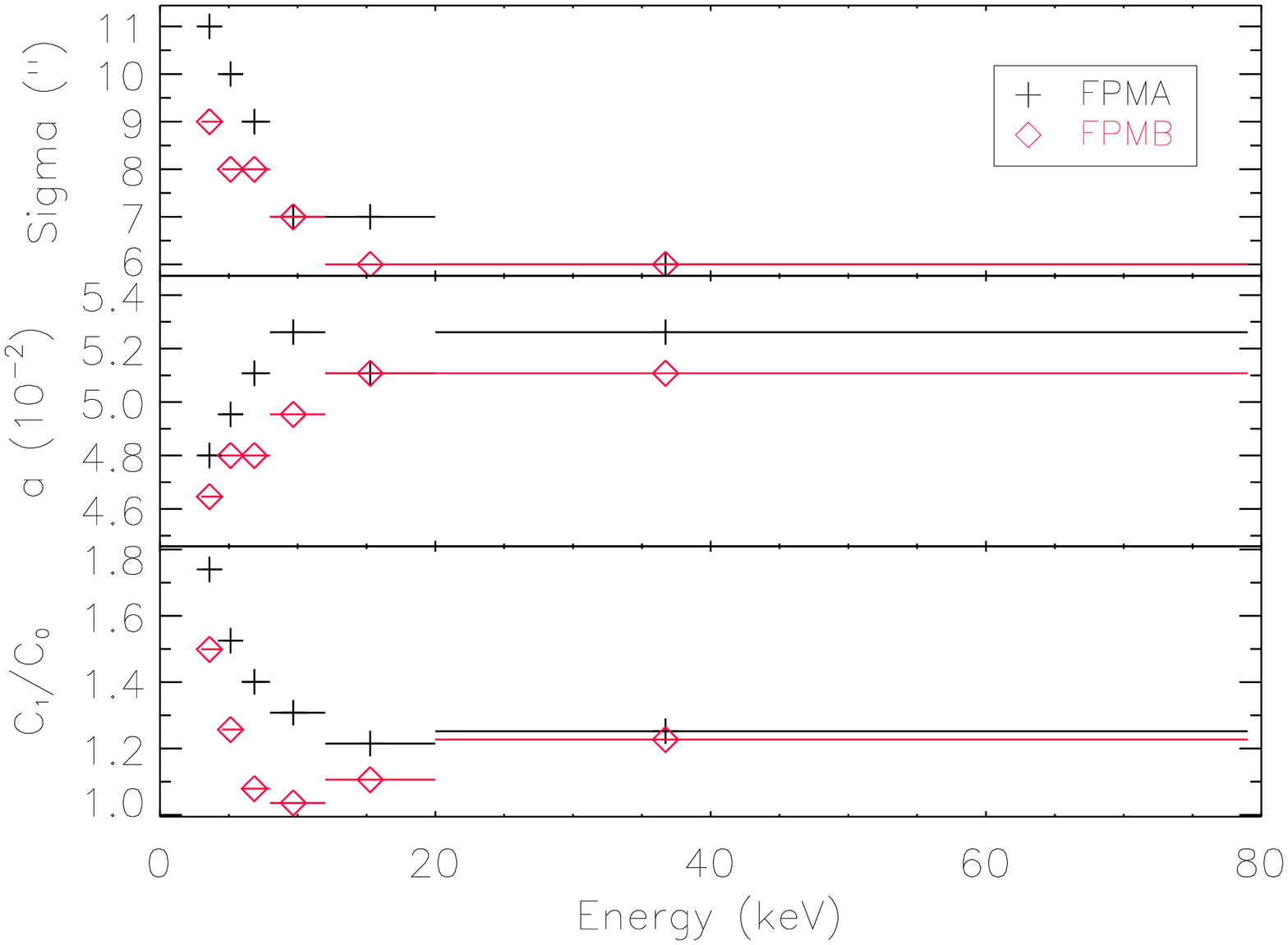} &
\hspace{-3.0 mm}
\includegraphics[width=2.45 in, angle=90]{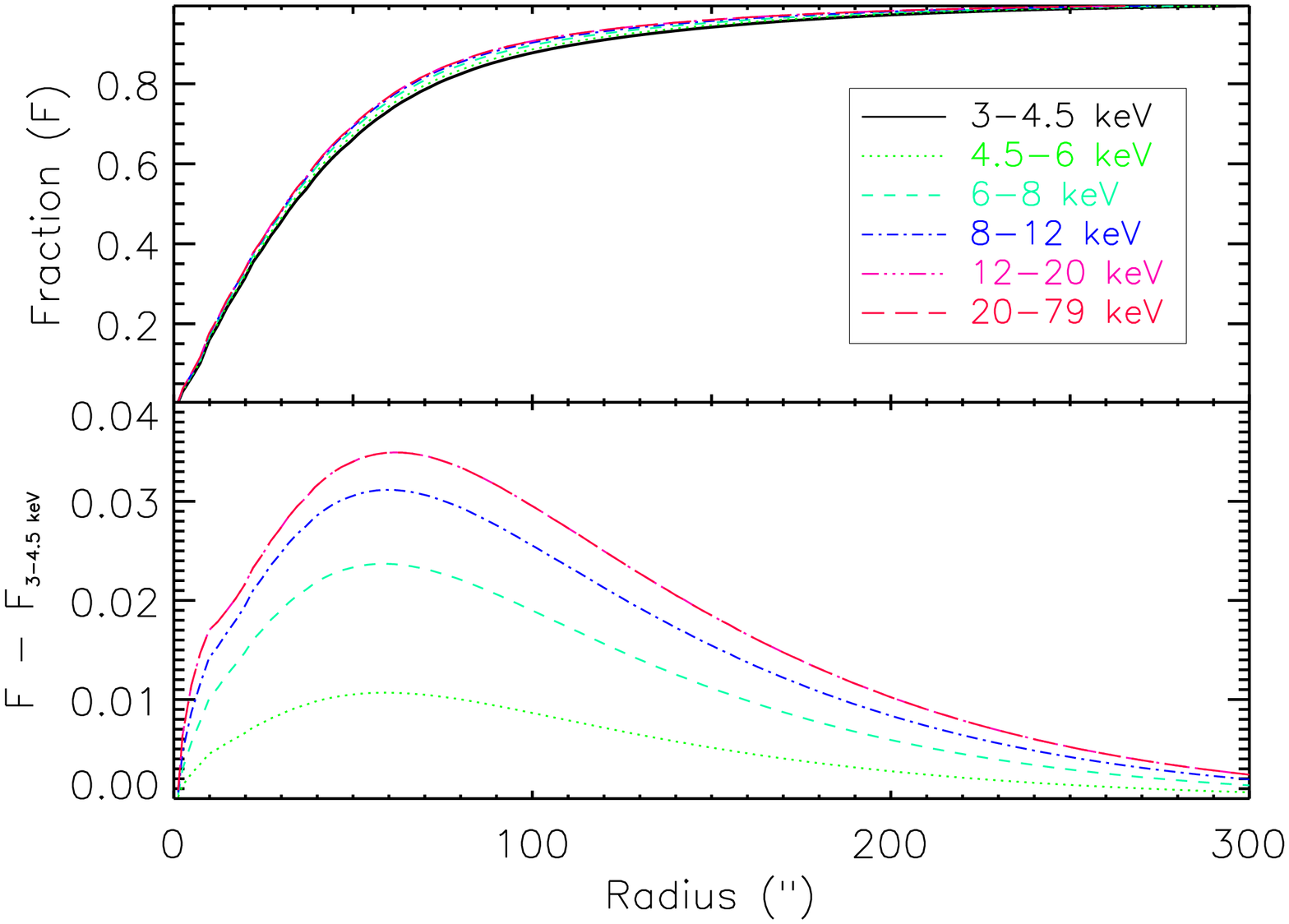} \\
\hspace{-15.0 mm}
\includegraphics[width=3.5 in]{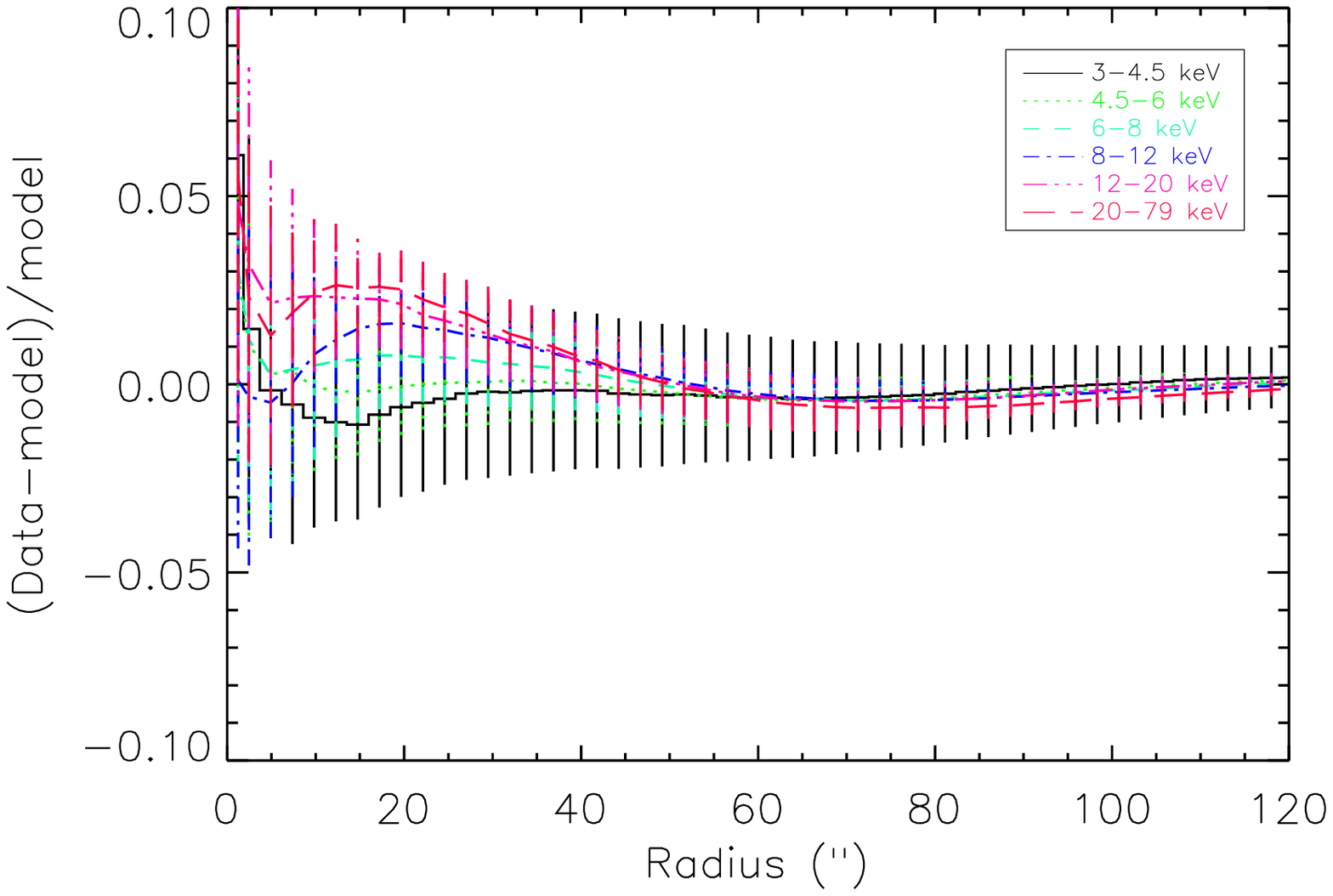} &
\hspace{-28.0 mm}
\includegraphics[width=3.5 in]{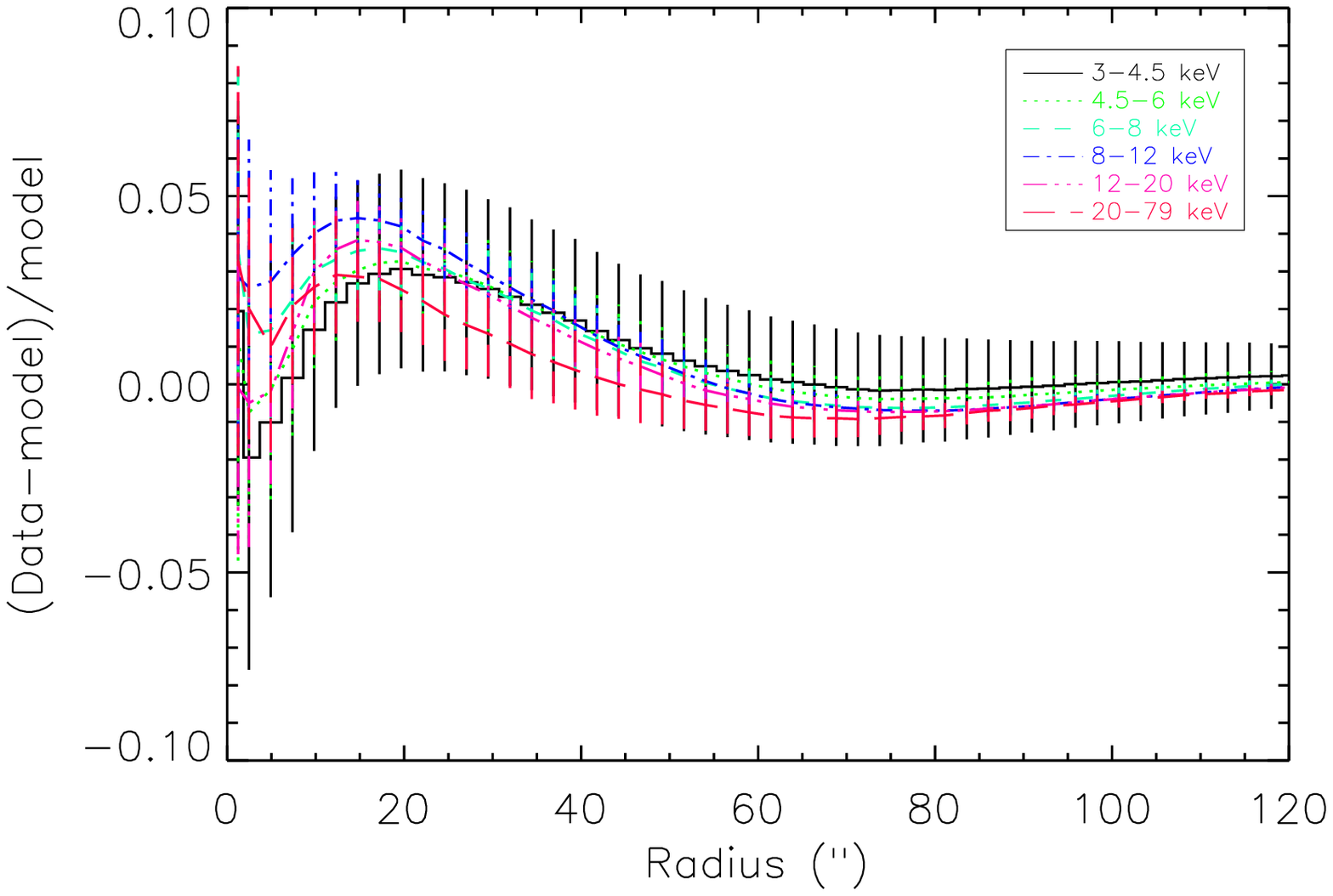} \\
\end{tabular}
\caption{{\it Top left}: Measured best-fit parameters for the new PSF model with energy
(see Eqs.~\ref{eq:psfG} and \ref{eq:psfwoff}).
{\it Top right}: EEFs in different energy bands (top) and the differences from
that in the 3--4.5 keV band (bottom) for FPMB.
{\it Bottom}: Differences between the EEFs of the nine near on-axis observations and that of
the PSF model for FPMA (left) and FPMB (right).
The solid lines are the average difference and the error bars are the standard deviations of the nine data sets.
\label{fig:parenc}
}
\vspace{2mm}
\end{figure}

The final model PSF includes all the modifications described above, the Gaussian convolution,
the exponential wing, and the off-axis corrections for the wing component. Using the final model PSF,
we fit the observed radial profiles of the near on-axis observations (see Section~\ref{sec:onpsf}),
and show the best-fit parameters for the Gaussian convolution width ($\sigma$),
the decay constant of the exponential wing ($a$), and
the relative normalizations ($C_{\rm 1}/C_{\rm 0}$) of the core ($C_{\rm 0}$)
and the wing ($C_{\rm 1}$) components for FPMA and FPMB in the top left panel of Fig.~\ref{fig:parenc}.

The results show that the PSF sharpens with energy; the model PSF requires smaller Gaussian width
and a narrower wing, as well as a relatively larger amplitude for the core component. 
This trend is seen for both FPMA and FPMB and saturates at $\sim$10 keV. 
The reason for this could be that the inner shells have slightly better surface profiles
than the outer shells \AR{do\cite{cab+11}. It is also possible that small contamination, for example,
by epoxy outgassing\cite{acd+09} might contribute to the energy dependence.}
The broadening and the wing correction to the ray-traced PSF are required in all the energy bands,
implying that there may be some energy independent effects as well (e.g., imperfect aspect correction
and/or mirror characterization). It is not possible to further speculate the origins of the broadening
and the wing with the empirical model presented in this work.
More complete studies using the ray-traced model and the observatory simulator (NuSIM) \cite{hlc+10} are required.

Since the PSF shape changes with energy, the EEF for an aperture will be different
for different energy bands. We measured the EEFs of the PSF model in different energy bands
and show them in top right of Figure~\ref{fig:parenc}. The difference is aperture dependent and is
maximum between the lowest and the highest energy bands for $\sim$60$''$ extraction.

Note that the best-fit parameters for the PSF model differ from observation to observation, while
the parameters in top left of Figure~\ref{fig:parenc} minimize the sum of the $\chi^2$ of the nine
near on-axis observations in Table~\ref{ta:psfobsid}.
Hence, using averaged PSF for different observations will produce errors for individual observation.
In order to estimate the errors,
we compared the EEF of the PSF model with those of the observations, calculated
the average and the standard deviation of difference between the model PSF and the data as a function
of radius, and show the results in Figure~\ref{fig:parenc} bottom. The difference is less than 5\% for
$R\gapp10''$ in any energy band, and $\lapp$3\% for $R\gapp 60''$, implying that the inaccuracy in the
effective area produced by the PSF model is $\lapp$3\% for extraction regions with $R\gapp 60''$.
Note that the energy dependence in the top-right plot of Figure~\ref{fig:parenc} is included
in the PSF model and no clear trend with energy is visible in the difference plots, and thus the PSF
model does not add any significant slope error to ARF for apertures of $R\gapp60''$.
However, we find systematic residuals for FPMB at $R$=10--20$''$ (Fig.~\ref{fig:parenc} bottom right),
which can be reduced by adding another component to the PSF model of FPMB.

In table~\ref{ta:hpdvals}, we show the average, standard deviation, minimum and maximum of
HPD measured with the nine near on-axis observations in different energy bands. Note that
those are observed HPD and may have been blurred by some effects such as imperfect aspect
reconstruction, and incomplete background modeling, and thus not the intrinsic instrumental HPD.
We find that HPD of the {\em NuSTAR} optics varies from $60''$ to $70''$ depending on the optics module
and the energy band, and that the PSF of FPMB has smaller HPD, which can also be inferred
from the PSF model parameters in top left of Figure~\ref{fig:parenc}; PSF at higher energy and of FPMB requires a sharper core.

\begin{table}[t]
\caption{Measured half power diameter for the on-axis observations in Table~\ref{ta:psfobsid}}
\center
\begin{tabular}{ccccc|cccc}
\hline
Energy & \multicolumn{4}{c|}{FPMA} & \multicolumn{4}{c}{FPMB} \\ \hline
 & HPD & $\sigma_{\rm HPD}$ & min. & max. & HPD & $\sigma_{\rm HPD}$ & min. & max.  \\
(keV) & ($''$) & ($''$) & ($''$) & ($''$)& ($''$) & ($''$) & ($''$) & ($''$) \\
\hline
\hline
3--4.5   & 70.3 & 2.4 & 66.7 & 75.5 & 65.6 & 2.4 & 62.9 & 69.6  \\
4.5--6   & 67.1 & 1.0 & 64.7 & 67.7 & 62.6 & 1.2 & 60.9 & 64.7  \\
6--8     & 64.7 & 1.0 & 62.8 & 65.7 & 60.7 & 1.4 & 58.8 & 63.7  \\
8--12    & 63.5 & 1.1 & 61.8 & 64.7 & 59.5 & 1.5 & 57.9 & 62.8  \\
12--20   & 63.4 & 1.1 & 61.8 & 64.7 & 60.3 & 1.2 & 58.8 & 62.8  \\
20--79   & 63.4 & 1.0 & 61.8 & 64.7 & 62.4 & 1.5 & 60.8 & 65.7  \\
\hline
\end{tabular}
\label{ta:hpdvals}
\end{table}

\medskip
\subsection{PSF Stability over Time}
\label{sec:psftime} 
\begin{figure}[b]
\centering
\begin{tabular}{cc}
\hspace{7.0 mm}
\includegraphics[width=2.5 in, angle=90]{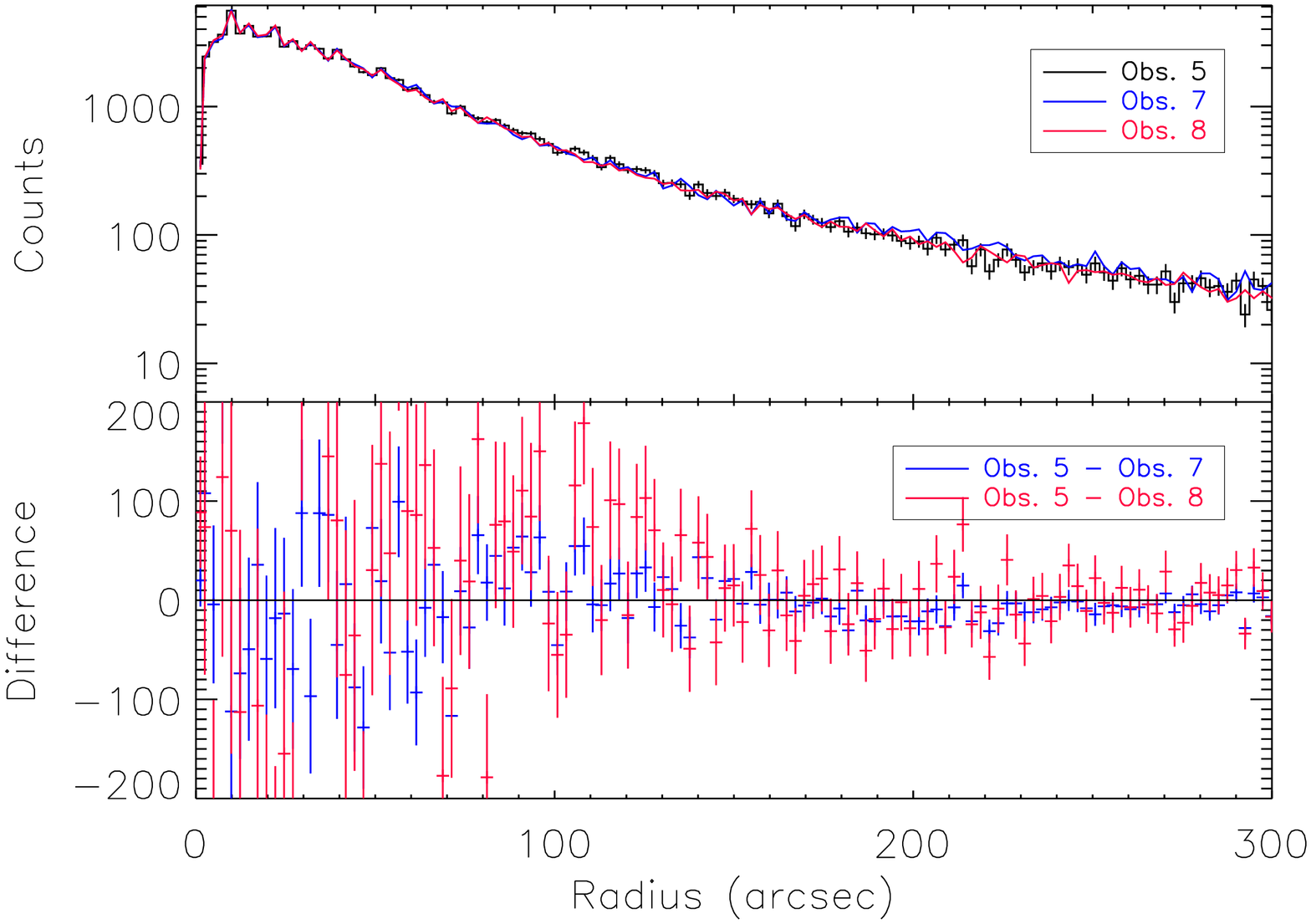} &
\hspace{-3.0 mm}
\includegraphics[width=2.5 in, angle=90]{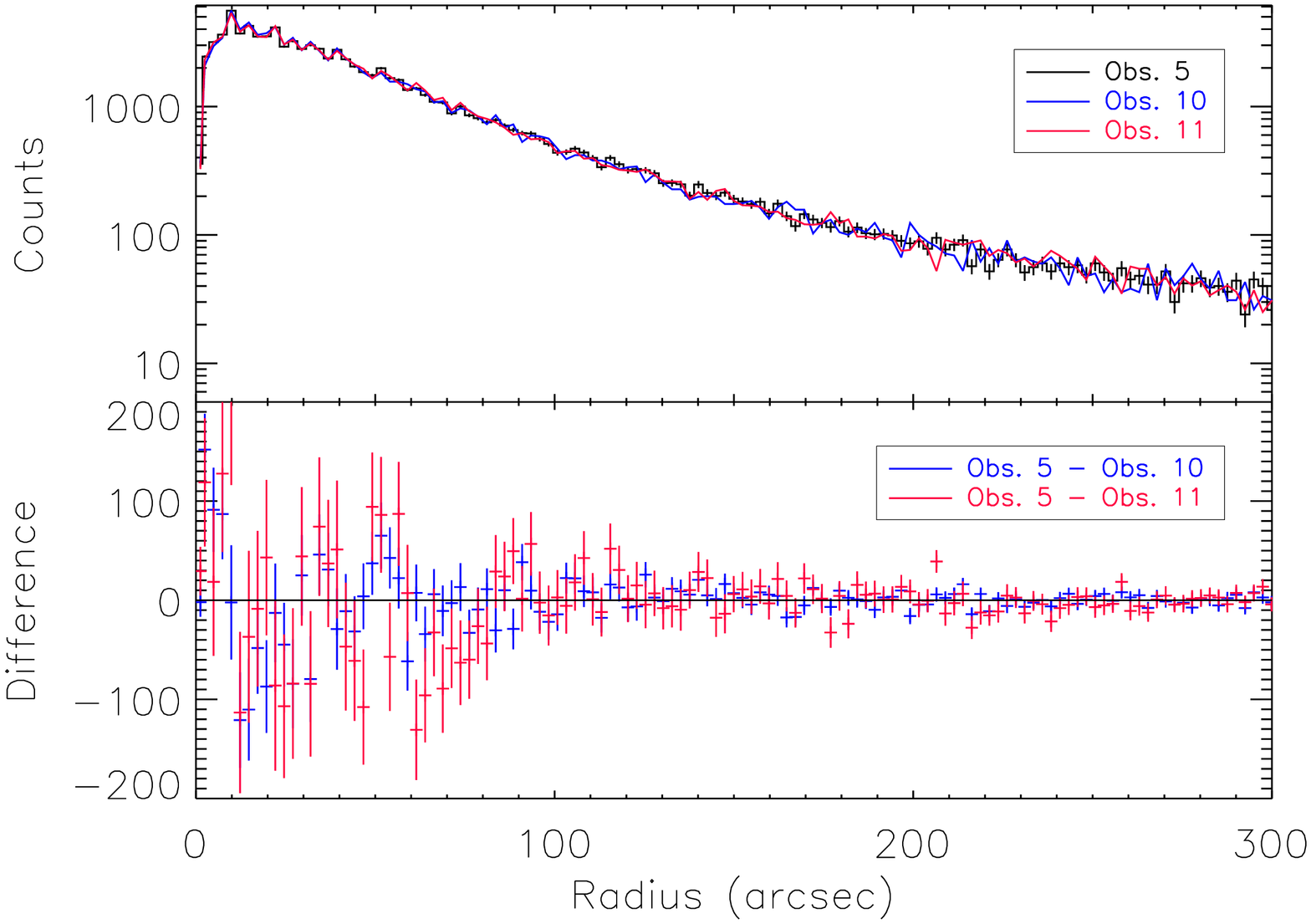} \\
\end{tabular}
\caption{{\it Left}: Observed FPMA radial profiles (top) for observations 5, 7 and 8 in the 3--4.5 keV band over $\sim$80 days and the differences
with 1$\sigma$ statistical-errors (bottom). {\it Right}: Same for observations 5, 10, and 11 over $\sim$300 days.
The profile at $T=0$ is shown in black, and those measured at later times are shown in blue and red.
\label{fig:PSFtime}}
\vspace{3mm}
\end{figure}

{\em NuSTAR} hard X-ray optics are composed of glass, graphite spacer and epoxy \cite{hab+10}. Although the materials
are carefully chosen so that the coefficients of thermal expansion match well with one another, the composite structure is
subject to change with temperature; for example, epoxy is known to creep with time by a thermal effect called
viscoelastic creep. In particular, temperature gradients applied to the optics by the Sun may have significant
impact on the optics structure. As the structure of the optics may change, the PSF may broaden over time.
In addition, outgassed epoxy molecules can stick to the mirror surface and scatter the incident X-rays.

Thermal effects and the epoxy outgassing on {\em NuSTAR} hard X-ray optics have been intensively
studied with ground experiments using prototype flat or Wolter-I optics.
The prototype optics underwent a thermal cycle or were exposed to large epoxy outgassing, and the surface profiles
and X-ray scattering properties of the optics were measured before and after the experiments, where
it was shown that the structure of {\em NuSTAR} optics will be stable at least over $\sim$10 years \cite{acd+09},
hence no significant change in PSF is expected over the period.
We verify this by comparing radial profiles of point source observations
(Table \ref{ta:psfobsid}) taken $\sim80$ days and $\sim300$ days apart.

We produced radial profiles of the observations 5, 7, 8, 10, and 11 (Table~\ref{ta:psfobsid}) in the energy bands used above.
In order to see any change in PSF, we subtracted the late-time profiles ($T \sim80$ and 300 days, observations 7, 8, 10 and 11)
from the reference profile ($T=0$, observation 5).
The radial profiles and the differences in the 3--4.5 keV are shown in Figure~\ref{fig:PSFtime},
which show no significant difference between the reference and the late-time profiles, considering difference
between profiles of two temporally adjacent observations.
The results are similar in the other energy bands as well. We therefore conclude that there was no significant change in the PSF.

\medskip
\section{Conclusion}
\label{sec:concl}
We calibrated the PSF of the {\em NuSTAR} hard X-ray optics using point source observations.
We show that the current ray-traced PSF alone is not able to describe the observed event distributions.
In order to explain the observed distributions, we convolved the ray-trace PSF with a Gaussian function,
added an extra wing component with small off-axis correction factors using empirical and analytic functions.
The modified PSF model describes the observations well,
and the error in the effective area produced by the PSF model is measured to be $\lapp$3\% for
extraction apertures with $R\gapp60''$. 
The PSF model described in this paper is included in the {\em NuSTAR} CALDB versions 20131007 and later.
We find that the PSF of the {\em NuSTAR} optics changes with energy, being sharper at higher energies.
Full ray-trace PSF modeling and {\em NuSTAR} observatory simulator need to be used for understanding
the energy dependent behavior. Finally, we show that angular response of {\em NuSTAR} optics has been stable over
the period of $\sim$300 days between 2012 July and 2013 April. Further studies will be conducted to improve
the accuracy of the PSF model and monitor the stability of the PSF.

\medskip
\acknowledgments     
This work was supported under NASA Contract No. NNG08FD60C, and  made use of data from the {\it NuSTAR} mission,
a project led by  the California Institute of Technology, managed by the Jet Propulsion  Laboratory,
and funded by the National Aeronautics and Space  Administration. We thank the {\it NuSTAR} Operations,
Software and  Calibration teams for support with the execution and analysis of  these observations.
This research has made use of the {\it NuSTAR}  Data Analysis Software (NuSTARDAS) jointly developed by
the ASI  Science Data Center (ASDC, Italy) and the California Institute of  Technology (USA). 


\medskip
\bibliography{ms}   

\begin{thebibliography}{10}

\bibitem{hcc+13}
F.~A. {Harrison}, W.~W. {Craig}, F.~E. {Christensen}, and et~al., ``{The
  Nuclear Spectroscopic Telescope Array (NuSTAR) High-energy X-Ray Mission},''
  {\em ApJ.}~{\bf 770}, p.~103, 2013.

\bibitem{mhb+14}
K.~K. {Madsen}, F.~A. {Harrison}, S.~E. {Boggs}, and et~al., ``{The Nuclear
  Spectroscopic Telescope Array (NuSTAR) High-energy X-ray Mission},'' {\em
  Proc. SPIE} {\bf 9144}, 2014.

\bibitem{sjs+95}
P.~J. {Serlemitsos}, L.~{Jalota}, Y.~{Soong}, and et~al., ``{The X-ray
  telescope on board ASCA},'' {\em PASJ.}~{\bf 47}, pp.~105--114, 1995.

\bibitem{hab+10}
C.~J. {Hailey}, H.~{An}, K.~L. {Blaedel}, and et~al., ``{The Nuclear
  Spectroscopic Telescope Array (NuSTAR): optics overview and current
  status},'' {\em Proc. SPIE} {\bf 7732}, p.~77320T, 2010.

\bibitem{z09}
W.~W. {Zhang}, ``{Manufacture of mirror glass substrates for the NuSTAR
  mission},'' {\em Proc. SPIE} {\bf 7437}, p.~74370N, 2009.

\bibitem{cjb+11}
F.~E. {Christensen}, A.~C. {Jakobsen}, N.~F. {Brejnholt}, and et~al.,
  ``{Coatings for the NuSTAR mission},'' {\em Proc. SPIE} {\bf 8147},
  p.~81470U, 2011.

\bibitem{cab+11}
W.~W. {Craig}, H.~{An}, K.~L. {Blaedel}, and et~al., ``{Fabrication of the
  NuSTAR flight optics},'' {\em Proc. SPIE} {\bf 8147}, p.~81470H, 2011.

\bibitem{bcj+11}
N.~F. {Brejnholt}, F.~E. {Christensen}, A.~C. {Jakobsen}, and et~al., ``{NuSTAR
  ground calibration: The Rainwater Memorial Calibration Facility (RaMCaF)},''
  {\em Proc. SPIE} {\bf 8147}, p.~81470I, 2011.

\bibitem{kab+11}
J.~E. {Koglin}, H.~{An}, N.~{Barri{\`e}re}, and et~al., ``{First results from
  the ground calibration of the NuSTAR flight optics},'' {\em Proc. SPIE} {\bf
  8147}, p.~81470J, 2011.

\bibitem{w11}
N.~J. Westergaard, ``{MT\_RAYOR}: A versatile raytracing tool for x-ray
  telescopes,'' {\em Proc. SPIE} {\bf 8147}, p.~1311, 2011.

\bibitem{whm+14}
D.~R. {Wik}, A.~{Hornstrup}, S.~{Molendi}, and et~al., ``{NuSTAR Observations
  of the Bullet Cluster: Constraints on Inverse Compton Emission},'' {\em ArXiv
  e-prints, ArXiv:1403.2722} , 2014.

\bibitem{vc72}
L.~P. {van Speybroeck} and R.~C. {Chase}, ``{Design parameters of
  paraboloid-hyperboloid telescopes for X-ray astronomy.},'' {\em Appl.
  Opt.}~{\bf 11}, pp.~440--445, 1972.

\bibitem{scbc+09}
D.~{Spiga}, V.~{Cotroneo}, S.~{Basso}, and P.~{Conconi}, ``{Analytical
  computation of the off-axis effective area of grazing incidence X-ray
  mirrors},'' {\em A\&A}~{\bf 505}, pp.~373--384, 2009.

\bibitem{acd+09}
H.~{An}, F.~E. {Christensen}, M.~{Doll}, and et~al., ``{Evaluation of epoxy for
  use on NuSTAR optics},'' {\em Proc. SPIE} {\bf 7437}, p.~74371J, 2009.

\bibitem{hlc+10}
D.~I. {Harp}, C.~C. {Liebe}, W.~{Craig}, F.~{Harrison}, K.~{Kruse-Madsen}, and
  A.~{Zoglauer}, ``{NuSTAR: system engineering and modeling challenges in
  pointing reconstruction for a deployable x-ray telescope},'' {\em Proc. SPIE}
  {\bf 7738}, p.~77380Z, 2010.

\end{thebibliography}
\bibliographystyle{spiebib}   

\end{document}